\documentclass[review, sort&compress]{elsarticle}

\usepackage{lineno,hyperref}
\usepackage{graphicx, epstopdf, amstext, amssymb,epsf,color,hyperref}

\modulolinenumbers[5]

\usepackage{amssymb}
\usepackage{graphicx}
\usepackage{epstopdf}
\usepackage{amsmath}
\usepackage{color}
\usepackage{verbatim}
\usepackage{cprotect}
\usepackage{hyperref}

\journal{Physica A}

\begin{document}

\begin{frontmatter}



\title{Modeling the dynamics of dissent}


\author[skku]{Eun Lee}

\author[skku,TIT]{Petter Holme}

\author[kias]{Sang Hoon Lee\corref{cor}\fnref{present}}
\ead{lshlj82@gntech.ac.kr}
\fntext[present]{Present address: Department of Liberal Arts, Gyeongnam National University of Science and Technology, Jinju 52725, Republic of Korea}

\cortext[cor]{Corresponding author}

\address[skku]{Department of Energy Science, 
Sungkyunkwan University, Suwon 16419, Republic of Korea}

\address[TIT]{Institute of Innovative Research, Tokyo Institute of Technology, Yokohama, Kanagawa 226--8503, Japan}

\address[kias]{School of Physics, Korea Institute for Advanced Study,
Seoul 02455, Republic of Korea}

\begin{abstract}
We investigate the formation of opinion against authority in an authoritarian society composed of agents with different levels of authority. We explore a ``dissenting'' opinion, held by lower-ranking, obedient, or less authoritative people, spreading in an environment of an ``affirmative'' opinion held by authoritative leaders. A real-world example would be a corrupt society where people revolt against such leaders, but it can be applied to more general situations. In our model, agents can change their opinion depending on their authority relative to their neighbors and their own confidence level. In addition, with a certain probability, agents can override the affirmative opinion to take the dissenting opinion of a neighbor. Based on analytic derivation and numerical simulations, we observe that both the network structure and heterogeneity in authority, and their correlation, significantly affect the possibility of the dissenting opinion to spread through the population. In particular, the dissenting opinion is suppressed when the authority distribution is very heterogeneous and there exists a positive correlation between the authority and the number of neighbors of people (degree). Except for such an extreme case, though, spreading of the dissenting opinion takes place when people have the tendency to override the authority to hold the dissenting opinion, but the dissenting opinion can take a long time to spread to the entire society, depending on the model parameters. We argue that the internal social structure of agents sets the scale of the time to reach consensus, based on the analysis of the underlying structural properties of opinion spreading.
\end{abstract}


\end{frontmatter}


\section{Introduction}
\label{sec:introduction}

How much an opinion against a firmly established authority can spread in a population is an important estimate of the population's adaptability~\cite{Sen1999}, in particular when there exists a strong heterogeneity in the distribution of influential power regarding opinion formation. The topic of opinion formation has been studied widely to reveal the hidden mechanisms of a collective opinion dynamics on social networks~\cite{Castellano2009,Galam1986,Galam2008,sociophys}. There has been a wide variety of opinion formation models, such as the voter model~\cite{Sood2005, Miguel2005}, the majority rule model~\cite{Galam2002}, the bounded confidence model~\cite{Deffuant2000}, and the Sznajd model~\cite{Sznajd2000}. 
Many opinion formation models have focused on the effect of heterogeneity in a network structure for a global consensus~\cite{Yang2009} and considered heterogeneous distributions of personal characteristics---gender, age, job, economic level, personal interests~\cite{Park2007,Fowler2008}, and so on, as those two areas of heterogeneity are important for opinion dynamics on networks~\cite{Aral2009,Ramos2015}. 

However, most of those opinion formation models mix the concept of heterogeneity in the individual level with the heterogeneity in social structures, even though structural and individual heterogeneity can be independent of each other~\cite{Eom2014,Hang2014}. 
Previous studies derive the personal heterogeneity in influential power from the structural heterogeneity, such as the number of neighbors (or ``degree,'' in the terminology of network science)~\cite{Yang2009} or PageRank (or ``eigenvector centrality'')~\cite{Kandiah2012,Chakhm2013}. 
There have been other attempts to highlight heterogeneity in individual attributes~\cite{Mobilia2015,Galam2013,Mauro2005,conviction_pre}, as well as in authority dispersion and in asymmetric options.~\cite{authority} and asymmetric opinions~\cite{dotheright}. Nevertheless, all these are different from the genuine authority dispersion, so we are still lack understanding of the transmission of opinions held by non-influential agents, grounded in the heterogeneity of both network-structural properties and influential power or authority. 
To address this need, in this paper, we investigate the following questions: when a population is composed of different levels of authority of agents, how can an opinion held by obedient agents with less influential power be spread to the whole population? How do the structure and authority collectively contribute to the spreading process? 
 
To answer these questions, we introduce a stylized opinion formation model in a population with the prescribed authority scores assigned to its agents, who are connected via networks~\cite{NewmanBook}. We assume heterogeneously distributed authority scores assigned to the agents, and each agent additionally has two essential characteristics: the willingness to uphold a dissenting opinion against authority and the confidence level for their own opinion. The probability of dissent is characterized by the parameter hinted at in the experiment of Milgram~\cite{Milgram1963}, which exemplifies the obedient tendency to an authoritative person's injustice order for the individual level, along with the tendency to resist the authority when there exist companions who would do so together. The agents apply the social comparison process to judge the relative authority level~\cite{Comparison2011}. 
An illustrative case is a corrupt society where authoritative agents have an agreement on a certain immoral decision, and a less influential population has a dissenting opinion against it. 
As the results of our analysis, we present the crucial role of the correlation between network structure and authority, via intrinsic social relations representing the authority comparison process.

\section{Model}
\label{sec:model}

To model the society presented in Sec.~\ref{sec:introduction}, for each individual we take heterogeneous degree distributions representing heterogeneous networks structures where individuals reside, heterogeneous authority levels of the individuals, and the correlation between them. 
In addition, we also incorporate individuals' inner characteristics for the confidence to their own opinion and willingness to follow the dissenting opinion. For heterogeneous structures, we construct a network composed of $N$ agents as nodes; thus, we use the terms ``agent'' and ``node'' interchangeably in this paper. The edges between the nodes represent the relationship between the nodes on which the authority comparison and the opinion spreading are based.

For network generation, we use an unweighted and undirected scale-free network (SFN) without self-loop and multiple edges, from the configuration model~\cite{Newman2001}. The degree distribution follows the power-law, $p(k) \sim k^{-\lambda}$, which yields a degree sequence $\{ k_i \}$ for node $i \in \{0, 1, \cdots, N-1 \}$ (thus there exist $N$ nodes in total). We set the minimum degree $k_{\mathrm{min}}=2$ for the initial network construction. To keep the overall connectivity, we use the largest connected component from the initially constructed network for the dynamics of our model. We verify that the change of network sizes in terms of the number of nodes and edges due to this selection process is negligible. 
Given the resultant connected network, we adjust the degree exponent $\lambda$ to control the degree heterogeneity, where the smaller $\lambda$ results in more heterogeneous degree distributions. When $\lambda = 2$, the average degree $\langle k \rangle \simeq 9$, and the maximum degree $k_{\mathrm{max}} \simeq 306$. For $\lambda = 3$, $\langle k \rangle \simeq 4$ and $k_{\mathrm{max}} \simeq 100$. As the control group compared to such heterogeneous structures, we also take the fully connected network to simulate the well-mixed population, which is expected to more accurately follow the result of the analytic derivation based on the mean-field approximation in Sec.~\ref{sec:Results}.

For the authoritarian structure, we assign an intrinsic authority score $\{ s_i \}$ to each node $i \in \{0, 1, \cdots, N-1 \}$. To generate heterogeneous authority scores, we extract random numbers (real numbers, in contrast to the natural numbers for the degree sequence $\{ k_i \}$ by definition) from the power-law distribution $p(s) \sim s^{-\gamma}$ with the minimum value of unity. The setup is inspired by the Pareto distribution~\cite{Lorenz1905} of wealth and income, which are indirect representatives of authority. Therefore, in general, we have two sets of power-law distributed values: $\lambda$ for the degrees $\{k_i\}$ and $\gamma$ for the authority scores $\{s_i\}$. For simplicity, however, we use the same power-law exponent for the authority score and degree, i.e., $\lambda = \gamma$ in our model, assuming that the same power-law exponent controls both structural and authoritarian heterogeneities. The set of authority scores $\{ s_i \}$ for agents $i \in \{0, 1, \cdots, N-1\}$ will be correlated with the agents' degree with different types of correlations. In addition, we take the two representative cases for the degree exponent to see the effect of the heterogeneity of degree distribution: relatively heterogeneous $\gamma = 2$ and relative homogeneous $\gamma = 3$. 
    
To investigate the effect of the correlation between authority and network structures~\cite{Eom2014,Hang2014}, we take three types of correlations: positive, no (uncorrelated), and negative correlations. The positive correlation implies that agents with higher authority scores have larger degree values. To control the correlation in practice, we sort both $\{ s_i \}$ and $\{ k_i \}$ from the smallest to the largest and match the indices of the sorted $\{s_i\}$ with the sorted $\{k_i\}$ in their exact order (as a result, the rank-based correlations such as Spearman's $r$ or Kendall's $\tau=1$). The negative correlation is achieved by the opposite way of ordering, i.e., matching the indices using the ascending order for $\{ s_i \}$ and the descending order for $\{ k_i \}$ (the rank-based correlations $= -1$). The uncorrelated case corresponds to the random matching (the rank-based correlations $\simeq 0$ on average).  
 
Each agent in the model selectively accepts her neighbor's opinion or keeps her own opinion, depending on the result of the authority comparison~\cite{Comparison2011}. In addition, she also has an intrinsically biased probability toward the dissenting opinion. We consider two personal characteristics for the comparison process: the confidence parameter $\alpha$ and the acceptance probability of the dissenting opinion $p$. Here, $\alpha$ and $p$ are global variables at a societal level, i.e., every agent has the same $\alpha$ and $p$, for simplification.  
Each agent $i$ has the time-dependent binary opinion variable ${\sigma}_i(t) \in \{ 0, 1 \}$ at time $t$, where $0$ represents the authoritative opinion and $1$ represents the dissenting opinion in our convention. If we assume a corrupt society where an immoral opinion of the dominant authority prevails, the dissenting opinion of less influential people would be a desirable choice. Throughout the paper, therefore, we use the expression ``dissenting'' or ``opposing'' opinion for the less influential people's opinion, as the opposite of the ``authoritative'' or ``affirmative'' opinion held by authoritative people.

\begin{table}[t]
\caption{
The decision table for ${\sigma}_i(t+1)$, where $\Theta[q]$ with $q \equiv q(s_i,s_j ; \alpha)$ in Eq.~(\ref{eq:impact}) is the Heaviside step function.
}
\begin{center}
\begin{tabular}{c|cc}
\hline
\hline
 & ${\sigma}_j (t) = 1$ & ${\sigma}_j (t) = 0$ \\
\hline
$p$ & $1$ & $ {\sigma}_i(t)\Theta[q]$ \\
\hline
$1-p$ & $1 - [1-{\sigma}_i(t)] \Theta[q]$  & ${\sigma}_i(t) \Theta[q]$ \\
\hline
\hline
\end{tabular}
\label{table:decision_simple} 
\end{center}
\end{table}

\begin{figure}
\begin{center}
\includegraphics[width=1.0\textwidth]{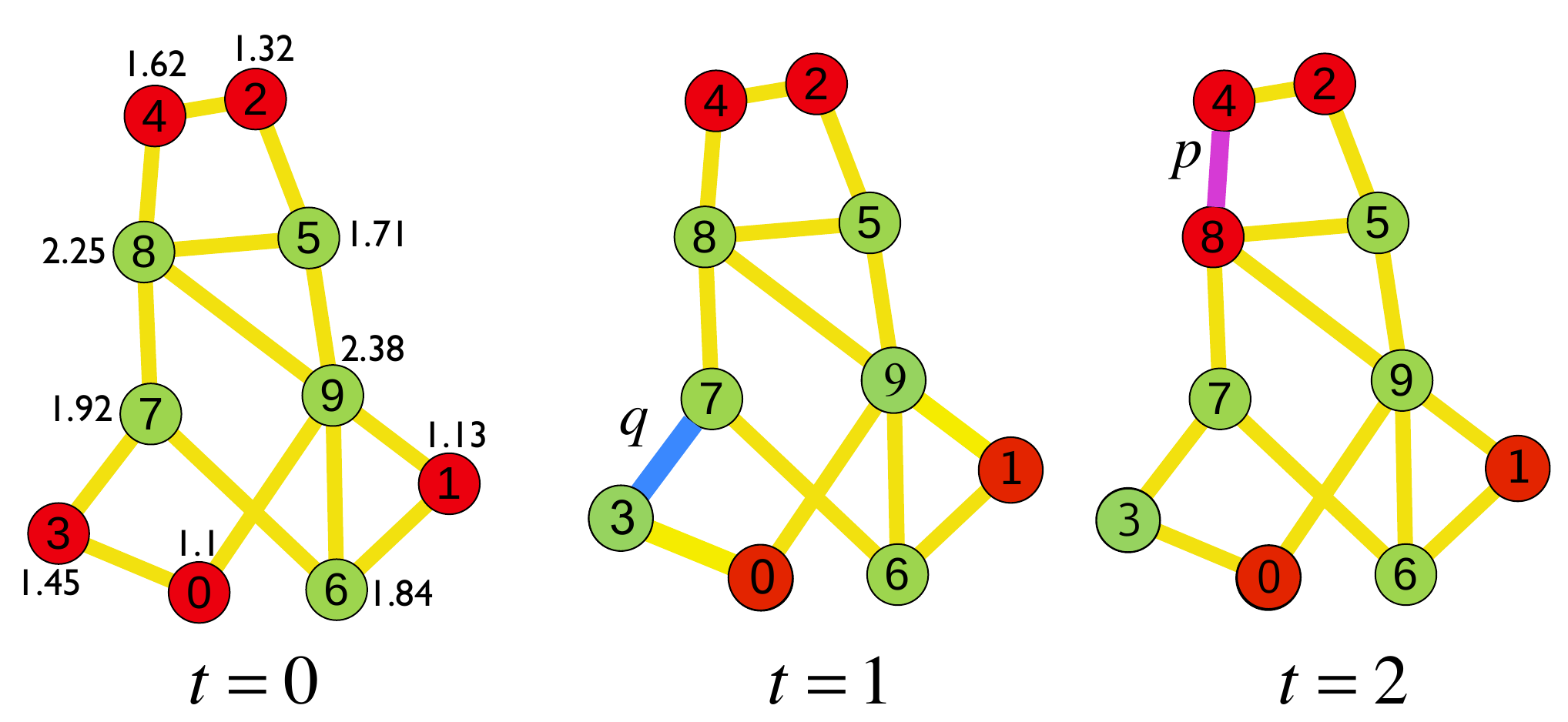}
\end{center}
\caption{Snapshots of the opinion change in our model. The numbers inside the nodes indicate the relative rank $i \in \{0, 1, \dots 9 \}$ for authority, where the larger values correspond to higher authority, which are used as the node indices in our description. The color of the nodes represents their opinion: red for the dissenting opinion ($\sigma_i = 1$) and green for the affirmative ($\sigma_i = 0$) opinions. The numbers outside the nodes indicate the authority scores $\{ s_i | i = 0, 1, \dots 9 \}$. The nodes compare their authority scores by Eq.~(\ref{eq:impact}). For instance, node $3$ chooses node $7$ for the comparison (the blue edge), and takes node $7$'s opinion ($\sigma_3 = 0 \to 1$ at $t=1$, because $\sigma_7(t=0) = 0$ and $q(s_3,s_7;\alpha) < 0$ (corresponding to the ``$q$'' process as denoted in the figure) in Eq.~(\ref{eq:impact}) when node $7$ has a larger authority score than that of node $3$. At $t = 2$, node $8$ chooses node $4$ (the purple edge), and with the probability $p$ (corresponding to the ``$p$'' process as denoted in the figure), regardless of their authority scores, node $8$ takes the dissenting ($\sigma_4 = 1$) opinion of node $4$ ($\sigma_8 = 0 \to 1$).
}
\label{fig:model_des}
\end{figure}

For the temporal evolution of the agents' opinion, at each time step, we select a node (denoted by $i$) uniformly at random and also choose one of its neighbors, denoted by $j$, uniformly at random. Then, node $i$ first checks node $j$'s opinion. If $\sigma_j (t) = 1$ (the dissenting opinion), with the probability $p$ (denoted by the ``$p$'' process in Fig.~\ref{fig:model_des}), she accepts the dissenting opinion of node $j$, i.e.,  ${\sigma}_i(t+1)=1$, regardless of her current opinion $\sigma_i(t)$ and their relative authority scores $s_i$ and $s_j$. With the complementary probability $1-p$ when $\sigma_j(t)=1$, or when ${\sigma}_j(t)=0$, node $i$ compares her authority score with that of node $j$ and decides whether or not she follows node $j$'s opinion, regardless of the current ${\sigma}_j (t)$ value (denoted by the ``$q$'' process in Fig.~\ref{fig:model_des}). The criterion is calculated based on their authority scores $s_i$ and $s_j$, and the confidence level $\alpha$. It is based on the impact function
\begin{equation}
q(s_i,s_j ; \alpha) \equiv \alpha s_i - (1-\alpha) s_j \,.
\label{eq:impact}
\end{equation}
If $q(s_i,s_j ; \alpha) \ge 0$, node $i$ keeps her opinion, i.e., ${\sigma}_i(t+1)={\sigma}_i(t)$. Otherwise, if $q(s_i,s_j ; \alpha) < 0$, node $i$ follows node $j$'s opinion, i.e., ${\sigma}_i(t+1) = {\sigma}_j (t)$. Therefore, large values of $\alpha$ represent a stronger tendency to keep the nodes' own opinion. We believe that this rule captures an aspect of human psychology revealed by the experiment of Milgram~\cite{Milgram1963}---the existence of a companion who raises the dissenting opinion is crucial for an individual's objection to the immoral authority. Note that the present model includes the conventional voter model as a limiting case when $p=\alpha=0$. Table~\ref{table:decision_simple} summarizes the rule, and Fig.~\ref{fig:model_des} illustrates an example case of the opinion evolution.

\section{Results}
\label{sec:Results}

In this section, we present the numerical simulation results supported by the analytic calculation on the stability condition of the dissenting opinion, in regard to the correlation between the degree and the authority score with different degree heterogeneity, namely, $\gamma = 2$ and $3$. We mainly focus on the final or steady-state fraction of the dissenting opinion in the network and the time to reach a consensus or steady state. 
The opinion averaged over agents and network realizations is expressed as
\begin{equation}
m(t) = \left\langle \sigma_{\nu;i}(t) \right\rangle = \frac{1}{n} \sum_{\nu=0}^{n-1} \left[ \frac{1}{N} \sum_{i=0}^{N-1} \sigma_{\nu;i}(t) \right] \,,
\label{m_avg}
\end{equation}
where $\nu \in \{0, 1, \cdots, n-1 \}$ is the index of independent realization of a network sample and $\sigma_{\nu;i}(t) \in \{0,1\}$ (recall that $0$ is the affirmative opinion and $1$ is the dissenting opinion) is the opinion of agent $i$ at time $t$, for the particular realization $\nu$. In the simulations, we take $N = 1000$ and $n = 2000$, unless otherwise stated.

The average opinion $m(t)$ eventually reaches the consensus $m(t) = 0$ or $m(t) = 1$ (which are the two absorbing states in our model, as no further change of opinion is possible once the network reaches one of the consensus states by the rule of our model) for a finite-size network, if we do not consider the practical time limitation. 
However, for finite-time simulations, there could be a steady state without reaching the consensus. When the average opinion $m(t)$ only slightly fluctuates around a specific finite value ($0 < m(t) < 1$) for a sufficient period, we consider the state as the balanced point for the opinion change from $0$ to $1$ and from $1$ to $0$, which we denote by the nontrivial steady state. 

We assume that the system reaches the nontrivial steady state if $m(t)$ fluctuates within a given range (denoted by $f$) for at least $t_c$ consecutive time steps, which is required for finite-size systems in finite-time numerical simulations. 
For practical simulations, we first wait for $t_{\max} = 1000$ time steps (we use the Monte Carlo time steps where $N$ trials of opinion changes correspond to a single unit of time, for numerical simulations) to check if the system reaches the absorbing states $m(t) = 0$ or $m(t) = 1$.
When $m(t < t_{\max}) = 0$ or $m(t < t_{\max}) = 1$, we halt the simulation and record the consensus time denoted by $\tau$.
If the system does not reach the absorbing states until $t = t_\mathrm{\max}$, we wait for the nontrivial steady state satisfying $| m_{\nu} (t-u+1) - m_{\nu} (t-u) | < f$ $\forall u \in \{0, 1, \cdots, t_c-2, t_c-1\}$ for $t_c$ consecutive times (we set $f = 0.05$ and $t_c = 1000$ based on the fluctuation in our observation). 
When $m(t)$ meets the nontrivial steady state criterion for the first time, we denote the time for reaching the nontrivial steady state by $\tau_s = t$. 
With $N = 1000$, the system always reaches the steady state with the given $f$ value, if it does not reach the consensus before $t_\mathrm{max}$. 
Note that the consensus states $m(t) = 0$ and $m(t) = 1$ are also (denoted by ``trivial,'' in that case) steady states, so $\tau_s = \tau$ for such cases. In other words, we denote both trivial and nontrivial steady states by $\tau_s$, and $\tau$ exclusively refers to the former case: consensus or absorbing states.
With this setting, we explore the equally spaced parameter ranges $\alpha \in \{0.00, 0.05, ..., 0.95, 1.00\}$ and $p \in \{0.00, 0.05, ..., 0.95, 1.00\}$.

\subsection{The fully connected network and the SFN without correlations}
\label{sec:fully_connected_case}

\begin{figure}
\begin{center}
\begin{tabular}{ll}
(a) & (b) \\
\includegraphics[width=0.4\textwidth]{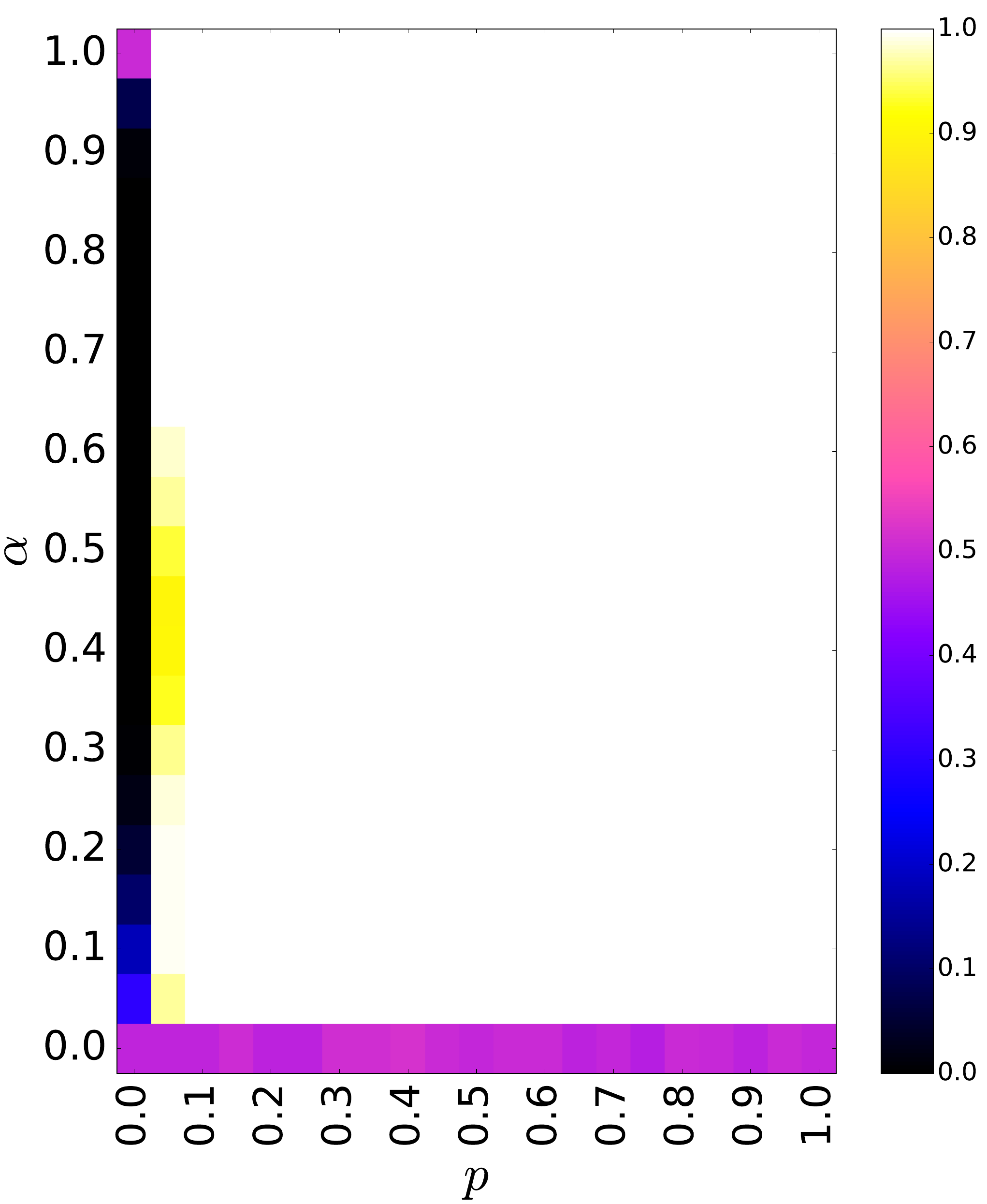} & \includegraphics[width=0.4\textwidth]{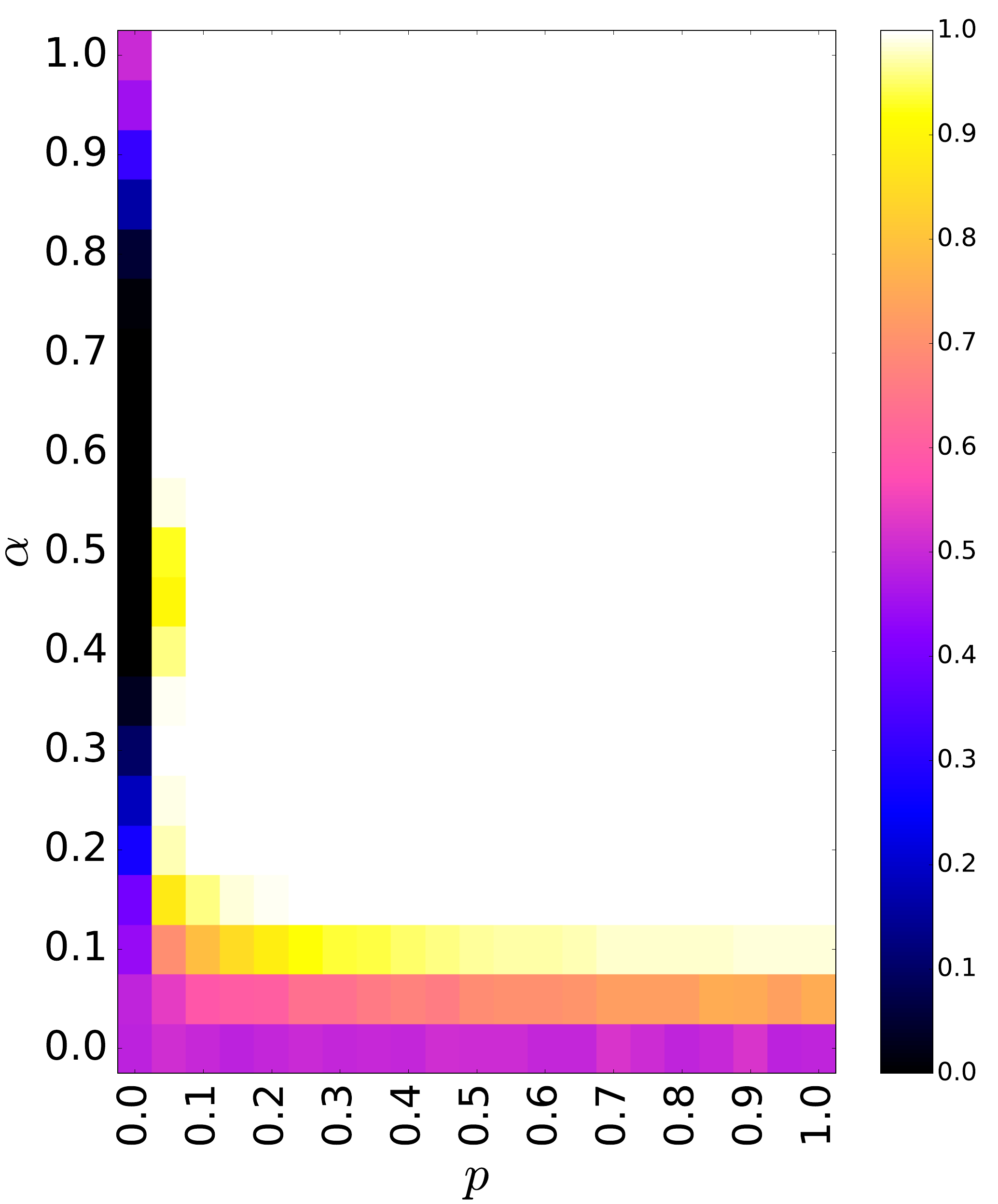} \\
\end{tabular}
\end{center}
\caption{The average opinion $m(\tau_s)$ at the steady state in Eq.~(\ref{m_avg}) with $N=1000$, averaged over $n = 2000$ realizations for the fully connected network case, with the authority heterogeneity exponents (a) $\gamma = 2$ and (b) $\gamma = 3$.}
\label{fig:fully}
\end{figure}

First, let us check the effects of authority heterogeneity only, by taking the fully connected network structure. Figure~\ref{fig:fully} shows $m(\tau_s)$ in the fully connected network in which the authority scores of individual nodes keep the unique heterogeneity. 
To understand the result analytically, 
we derive the stability condition for the steady state of opinions. Basically, at time $t$, node $i$ [with the authority score $s_i$ and opinion $\sigma_i(t)$] interacts with its random neighbor $j$ [with $s_j$ and $\sigma_j(t)$]. Then, node $i$'s opinion at time $t+1$ is determined as Table~\ref{table:decision_simple}. 

In this case, all of the agents are statistically equivalent and the neighbors are chosen uniformly at random (well-mixed population). If we denote the fraction of agents with the opinion $1$ at time $t$ by $m(t)$, the probability of $\sigma_i(t) = 1$ and that of $\sigma_j(t) = 1$ are also $m(t) = \langle \sigma_i(t) \rangle$, where the angular bracket denotes the agent-and-ensemble-averaged quantity and $m(t)$ becomes equivalent to the definition in Eq.~(\ref{m_avg}). According to Table~\ref{table:decision_simple}, with the shorthand notation $q \equiv q(s_i,s_j ; \alpha)$ in Eq.~(\ref{eq:impact}) and $m \equiv m(t)$, the probability of $\sigma_i(t+1) = 1$, or equivalently the average opinion of $i$ is
\begin{equation}
\begin{aligned}
\langle \sigma_i (t+1) \rangle = & \,\, p m + p (1 - m) m \mathrm{Pr}[ q \ge 0 ] \\
& + (1-p)m\{m + (1-\mathrm{Pr}[ q \ge 0 ] - m (1-\mathrm{Pr}[ q \ge 0 ])\} \\
& + (1-p) (1 - m) m \mathrm{Pr}[ q \ge 0 ]\,,
\label{eq:stability_condition_formula}
\end{aligned}
\end{equation}
where we assume the independence of the current opinion and (static) authority and $\mathrm{Pr}[q \ge 0]$ denotes the probability that the inequality $q \ge 0$ holds. 
Rearranging all of the terms and imposing the steady state condition $\langle \sigma_i(t+1) \rangle = \langle \sigma_i(t) \rangle = m$, we obtain 
\begin{equation}
p m \mathrm{Pr}[ q \ge 0 ] = p m^2 \mathrm{Pr}[ q \ge 0 ] \,,
\label{eq:step_function_condition}
\end{equation}
where replacing the instantaneous opinions $\sigma_i(t)$ and $\sigma_i(t+1)$ with the averaged opinions $\langle \sigma_i(t) \rangle = m(t)$ 
and $\langle \sigma_i(t+1) \rangle = m(t+1)$ corresponds to our mean-field assumption.

\begin{figure}
\begin{center}
\begin{tabular}{l}
(a) \\
\includegraphics[width=0.5\textwidth]{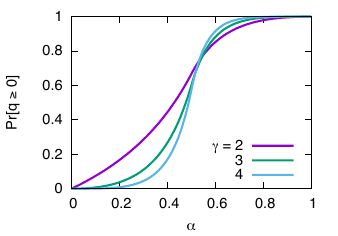} \\
(b) \\
\includegraphics[width=0.5\textwidth]{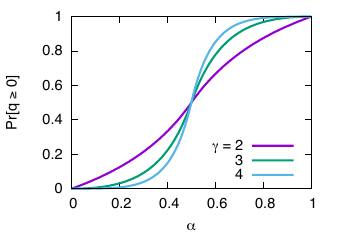} \\
(c) \\
\includegraphics[width=0.5\textwidth]{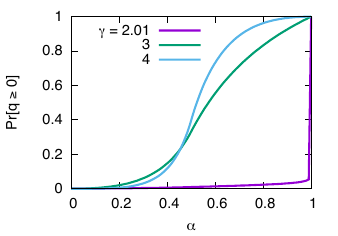}
\end{tabular}
\end{center}
\caption{$\mathrm{Pr}[ q \ge 0 ]$ for (a) the negatively correlated network case [Eq.~(\ref{eq:general_gamma_alpha})], (b) uncorrelated network case [Eq.~(\ref{eq:general_gamma_alpha_correlated})], and (c) positively correlated network case [Eq.~(\ref{eq:general_gamma_alpha_negative_correlated})].
}
\label{fig:Pr}
\end{figure}

For the network without any correlation between the degree and the authority score as the simplest case, which corresponds to both the fully connected network and the SFN without any correlation between the degree and the authority scores, let us consider the explicit form of $\mathrm{Pr}[ q \ge 0 ]$.
We give the power-law form of the authority distribution
$p(s) = (\gamma-1) s^{-\gamma}$ with $s_\mathrm{min} = 1$ [for the proper normalization
$\int_1^{\infty} ds \: p(s) = 1$]. Then, because $p(s_i)$ and $p(s_j)$ are independent to each other, we express $\mathrm{Pr}[q\ge 0]$ as
\begin{equation}
\begin{aligned}
\mathrm{Pr}[ q \ge 0 ] & = \int_1^{\infty} ds_j \int_1^{\infty} ds_i\: 
p(s_i)\: p(s_j) \Theta[q] \\
 & = \int_1^{\infty} ds_j \:p(s_j) \int_{\max{[1,(1-\alpha)s_j / \alpha]}}^{\infty} ds_i\:
p(s_i) \,,
\end{aligned}
\label{eq:Pr_integral}
\end{equation}
where $\Theta(q)$ is the Heaviside step function ($=1$ when $q \ge 0$ and $=0$ when $q < 0$).
When $\alpha \le 1/2$ [thus $(1-\alpha)/\alpha \ge 1$], 
\begin{equation}
\max\left( 1, \frac{1-\alpha}{\alpha} s_j \right) = \frac{1-\alpha}{\alpha} s_j \,,
\label{eq:alpha_small_condition}
\end{equation}
always, as $s_i \ge 1$.
Therefore, the integral in Eq.~(\ref{eq:Pr_integral}) becomes
\begin{equation}
\displaystyle \int_1^{\infty} ds_j\: p(s_j) \int_{(1-\alpha)s_j / \alpha}^{\infty} ds_i \:p(s_i) \,.
\label{eq:integral_not_split}
\end{equation}
When $\alpha > 1/2$ [thus $(1-\alpha)/\alpha < 1$],
\begin{equation}
\max\left( 1, \frac{1-\alpha}{\alpha} s_j \right) = 
\begin{cases}
\displaystyle \frac{1-\alpha}{\alpha} s_j , & \textrm{for } \displaystyle s_j \ge \frac{\alpha}{1-\alpha}\\
\displaystyle 1 , & \textrm{for } \displaystyle s_j < \frac{\alpha}{1-\alpha} \,,
\end{cases}
\label{eq:alpha_large_condition}
\end{equation}
so we have to split the integration range for $s_j$ in Eq.~(\ref{eq:Pr_integral}) as
\begin{equation}
\begin{array}{l}
\displaystyle \int_1^{\infty} ds_j \:p(s_j) \int_{\max{\{1,(1-\alpha)s_j / \alpha\}}}^{\infty} ds_i\: p(s_i) \\
\\
 = \displaystyle \int_1^{\alpha/(1-\alpha)} ds_j \:p(s_j) \int_1^{\infty} ds_i \:p(s_i) \\
\\
 + \displaystyle \int_{\alpha/(1-\alpha)}^{\infty} ds_j \:p(s_j) \int_{(1-\alpha)s_j/\alpha}^{\infty} ds_i \:p(s_i) \,. 
\end{array}
\label{eq:integral_split}
\end{equation}
Combining the two cases, we obtain
\begin{equation}
\mathrm{Pr}[ q \ge 0 ] = 
\begin{cases}
\displaystyle \frac{1}{2} \left( \frac{\alpha}{1-\alpha} \right)^{\gamma-1} , & \textrm{for $\alpha \le 1/2$} \\
\displaystyle 1 - \frac{1}{2} \left( \frac{1-\alpha}{\alpha} \right)^{\gamma-1} , & \textrm{for $\alpha > 1/2$} \,. 
\end{cases}
\label{eq:general_gamma_alpha}
\end{equation}
    When $\alpha = 1/2$, Eq.~(\ref{eq:general_gamma_alpha}) gives $\mathrm{Pr}[ q \ge 0 ] = 1/2$ guaranteeing the continuity of the function. Another limiting case is $\alpha=1$, where $\mathrm{Pr}[ q \ge 0 ] = 1$ (agent $i$ always beats agent $j$).
Figure~\ref{fig:Pr}(a) shows the functional form of $\mathrm{Pr}[ q \ge 0 ]$ given by Eq.~(\ref{eq:general_gamma_alpha}).
If we substitute $\mathrm{Pr}[ q \ge 0 ]$ in Eq.~(\ref{eq:general_gamma_alpha}) into Eq.~(\ref{eq:step_function_condition}), as we usually consider $\gamma > 1$, the steady-state with $0 < m < 1$ is possible for
$\alpha = 0$ or $p = 0$.

\begin{figure}
\begin{center}
\begin{tabular}{lll}
(a) & (b) & (c) \\
\includegraphics[width=0.3\textwidth]{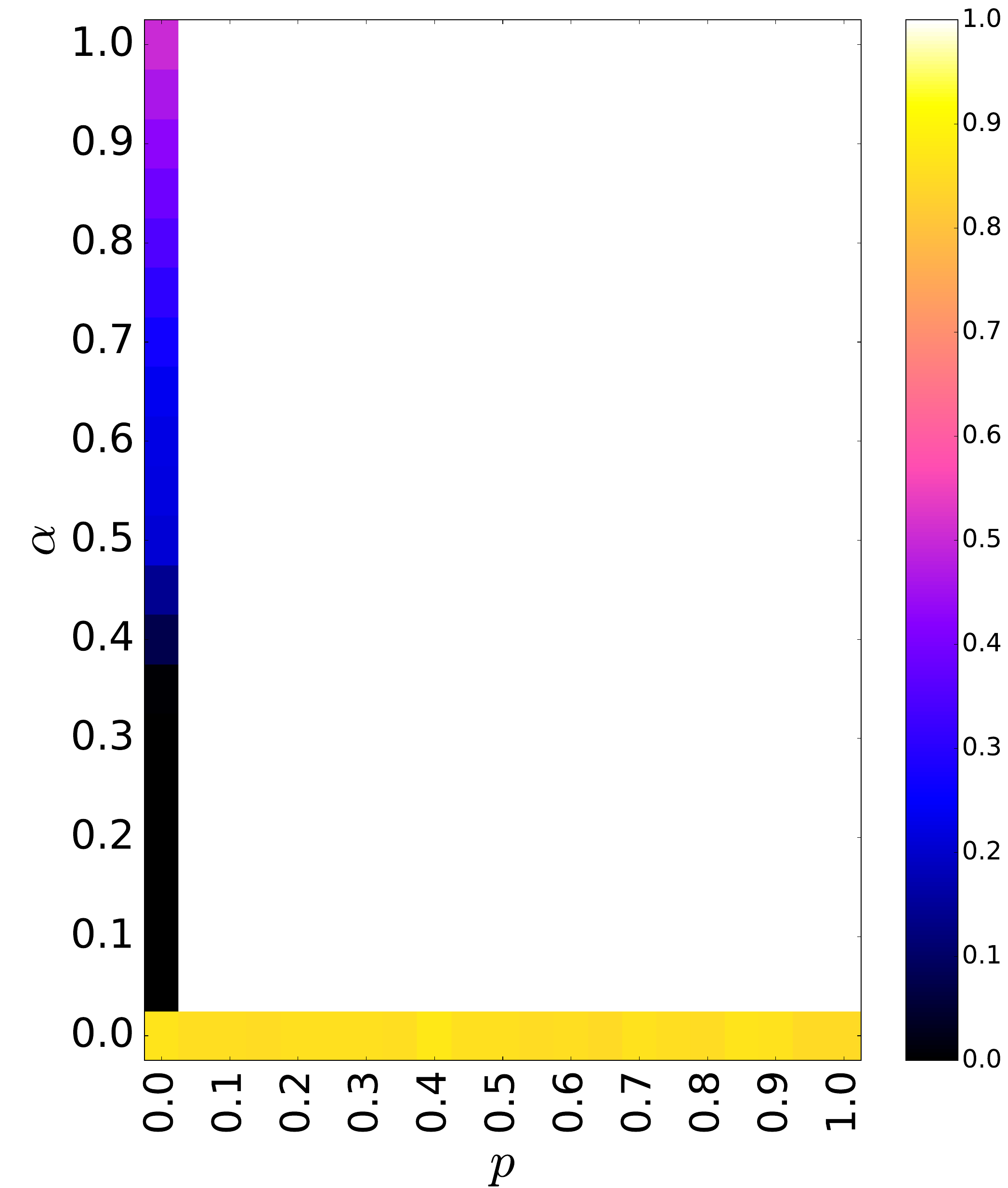} &
\includegraphics[width=0.3\textwidth]{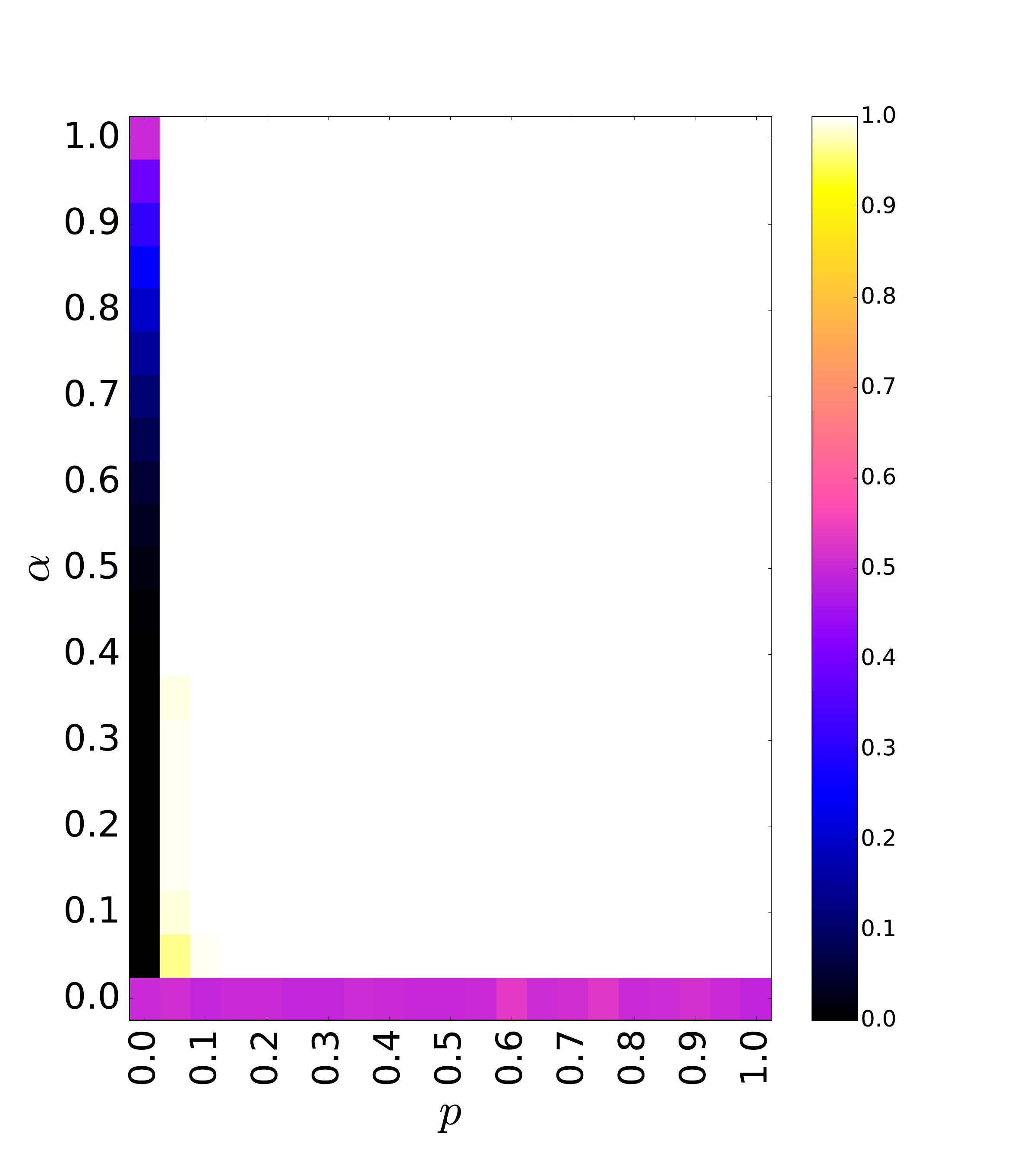} &
\includegraphics[width=0.3\textwidth]{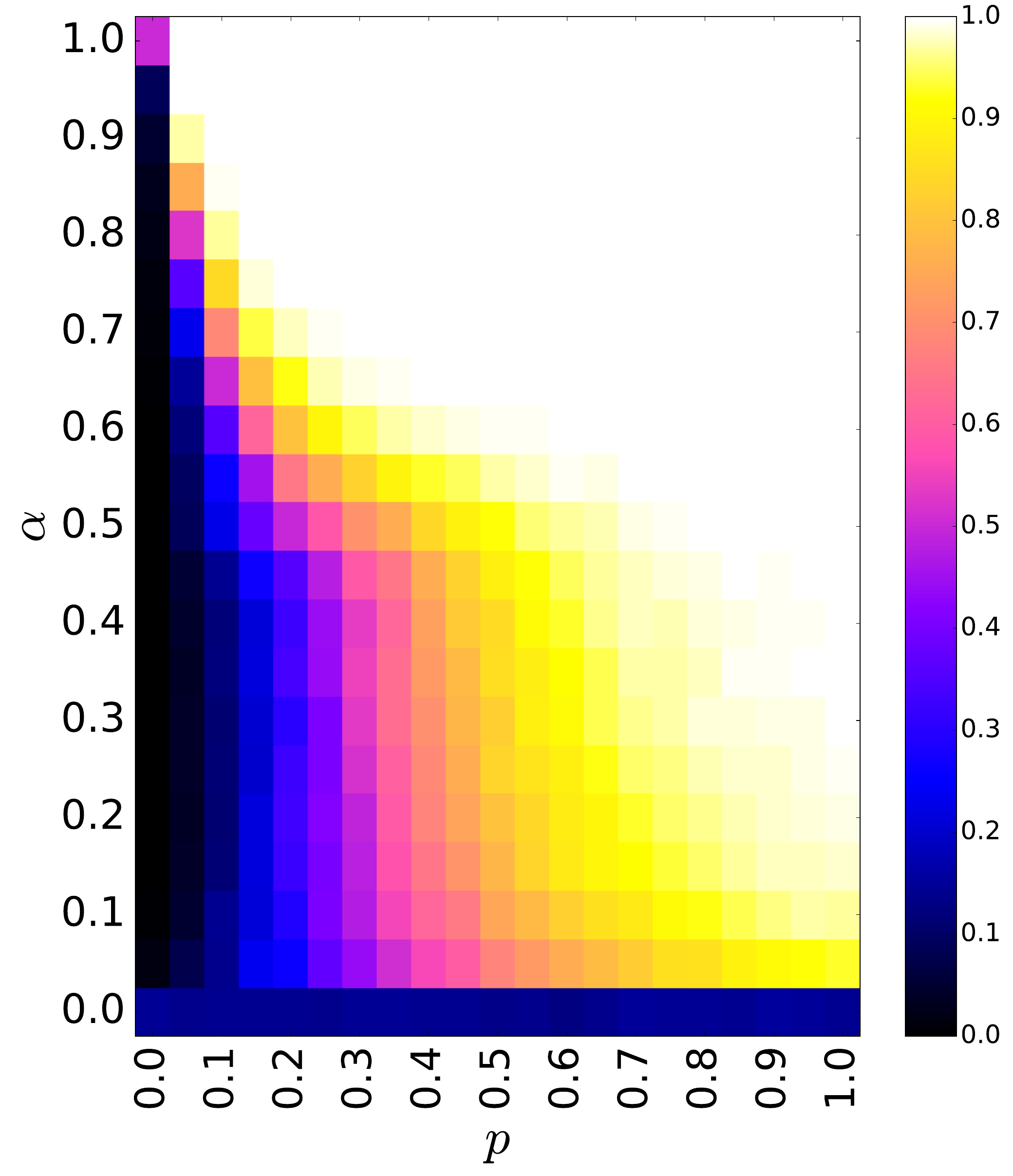}  \\
\end{tabular}
\begin{tabular}{lll}
(d) & (e) & (f) \\
\includegraphics[width=0.3\textwidth]{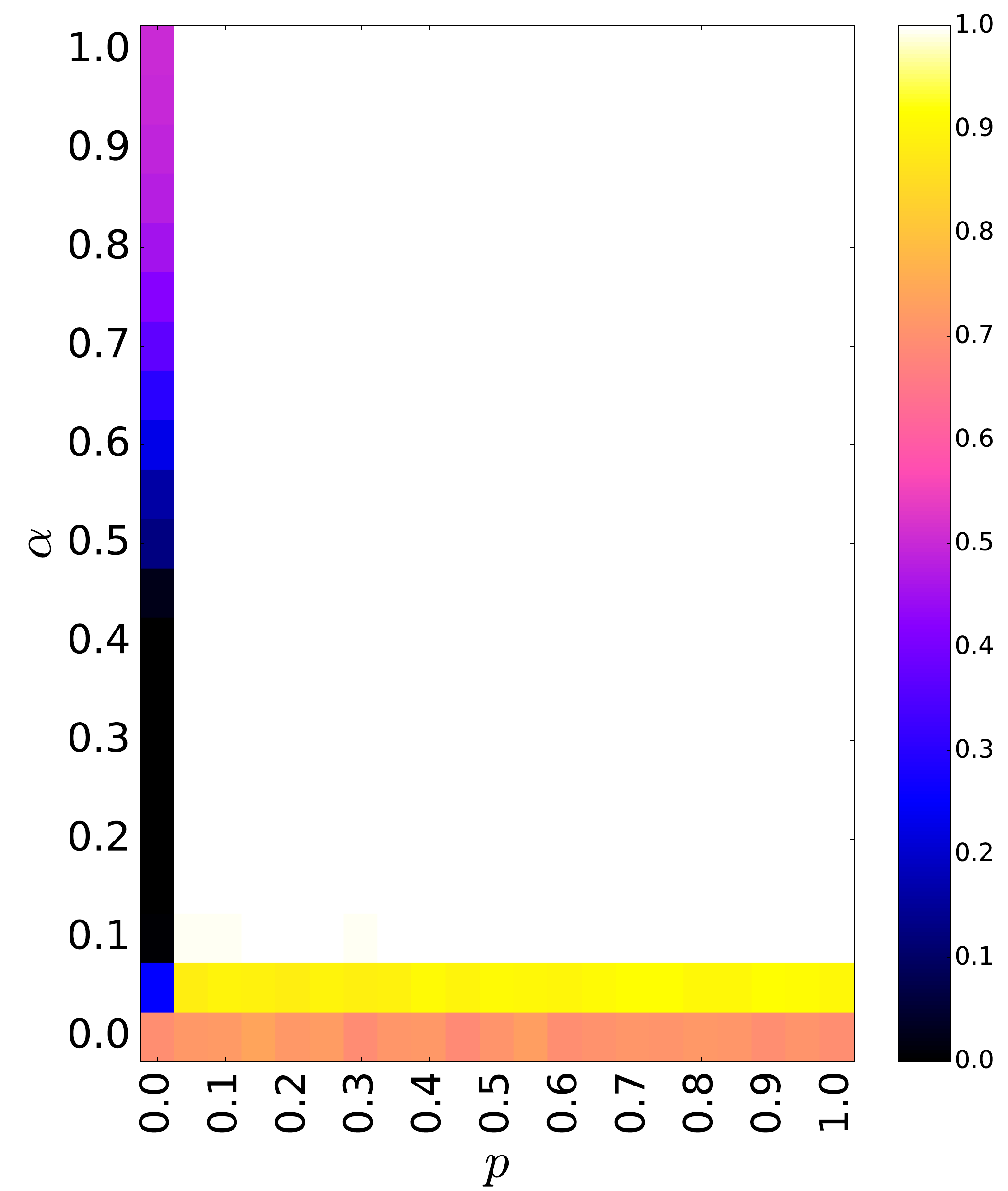} &
\includegraphics[width=0.3\textwidth]{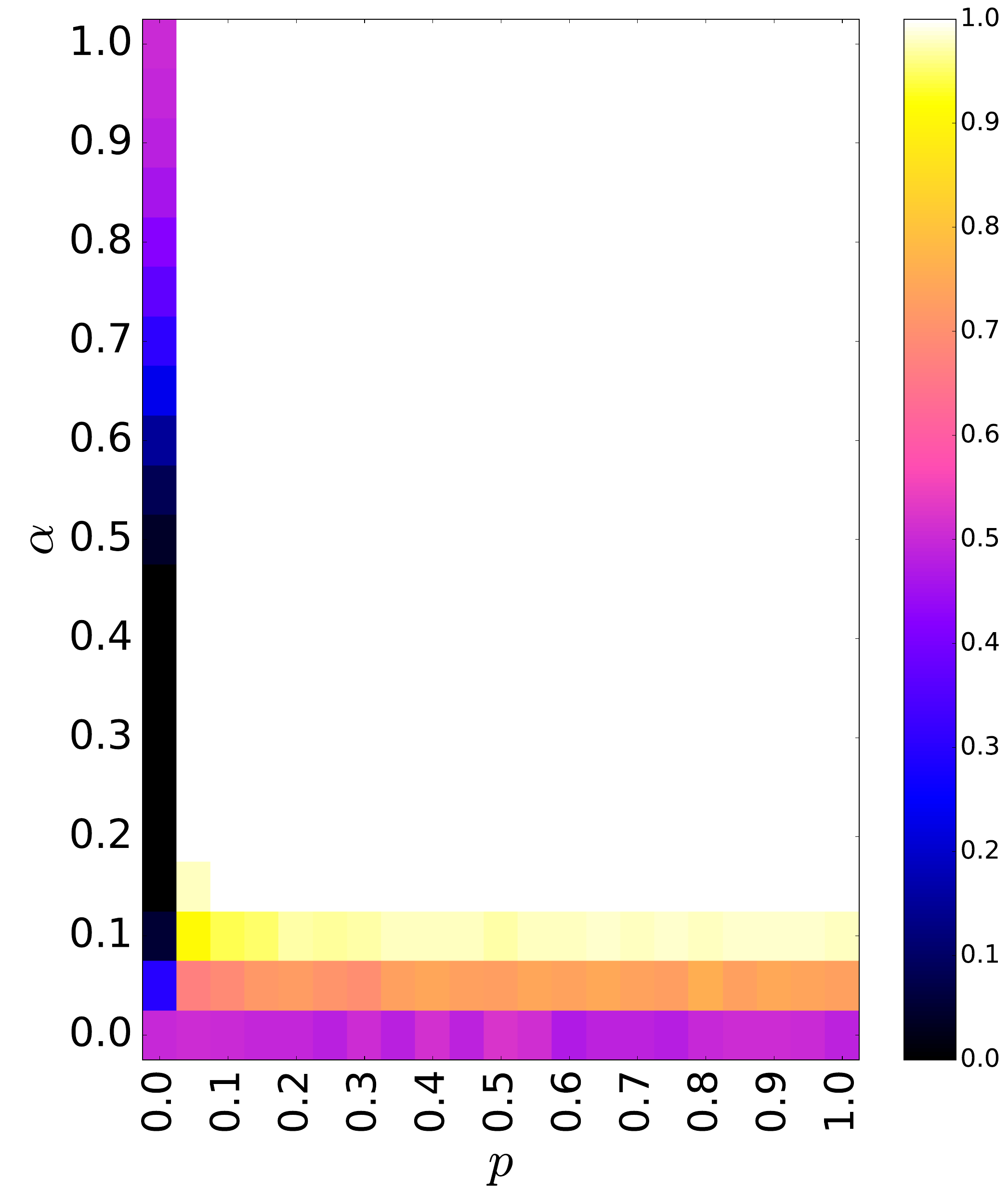} &
\includegraphics[width=0.3\textwidth]{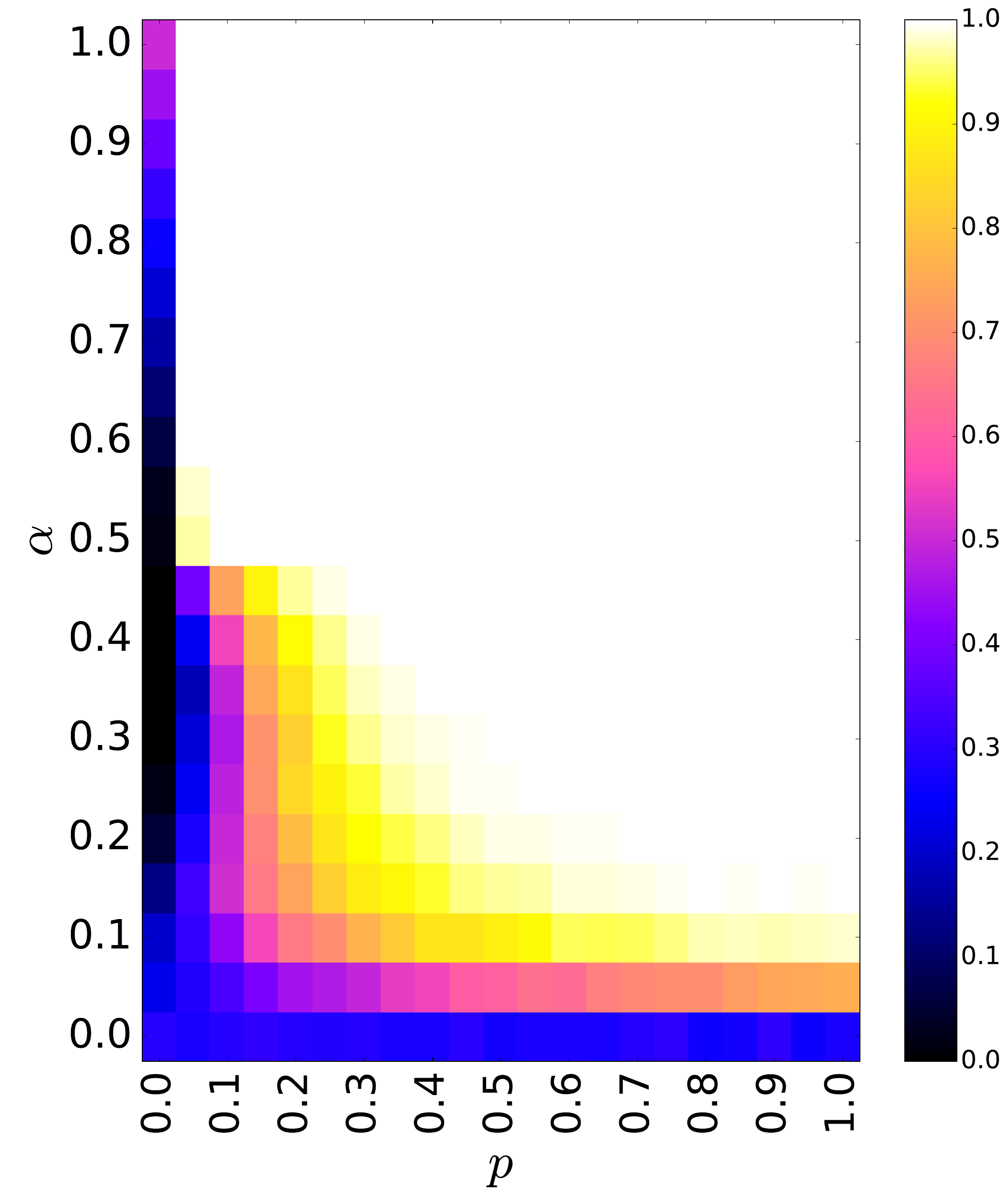}  \\
\end{tabular}
\end{center}
\caption{The average opinion $m(\tau_s)$ at the steady state in Eq.~(\ref{m_avg}) with $N=1000$, averaged over $n = 2000$ realizations with different correlations between degree $k$ and authority score $s$ in different power-law exponents for authority and degree distribution. The upper panels are the $\gamma = 2$ cases with (a) negative, (b) no, and (c) positive correlations. The lower panels are the $\gamma = 3$ cases with (d) negative, (e) no, and (f) positive correlations.}
\label{fig:gamma}
\end{figure}
	
Therefore, when $p > 0$ and $\mathrm{Pr}[ q \ge 0 ] > 0$, for the system to reach the steady state, $m$ should be either $0$ or $1$ (in practice, due to the intrinsic asymmetry between $0$ and $1$ for $p > 0$, the simulation results almost always converge to $m = 1$) or the consensus. The nontrivial steady-state with
$0 < m < 1$ is possible only for $\mathrm{Pr}[ q \ge 0 ] = 0$ or $p = 0$. 
From the results, we confirm the clear $\mathsf{L}$-shaped boundary on $\alpha = 0, p =0$ axes, and it matches well with the numerical result shown in Fig.~\ref{fig:fully}.
In the $\alpha > 0$ and $p > 0$ regime, there is a successful transmission of the dissenting opinion initially held by lower-ranked nodes in terms of authority. 
The result $m(\tau_s) \simeq 0.5$ at $\alpha = p=0$ is also consistent with the conventional voter model at that point. 
The only difference between $\gamma = 2$ and $\gamma = 3$ (Fig.~\ref{fig:fully}) is a larger strap for nonzero $m$ values near the $p=0$ axis for the $\gamma =3$ case [Fig.~\ref{fig:fully}(b)] than that of the $\gamma = 2$ case [Fig.~\ref{fig:fully}(a)]. 

Again, we would like to emphasize that the analytic derivation up to this point applies not only to the fully connected network, but also to the SFN without any correlation between the degrees and authority scores, as shown in Figs.~\ref{fig:gamma}(b) and \ref{fig:gamma}(e), because the probability of choosing node $j$ (proportional to her degree $k_j$~\cite{Eom2014,Hang2014}) is independent of her authority score $s_j$ for both cases. To be more precise, all of the elements of $\{ k_j | j = 0, 1, \cdots, N-1 \}$ themselves are identical for the fully connected network, and $k_j$ is independent of $s_j$ for the SFN without any correlation between $k_j$ and $s_j$. Therefore, as expected, the same $\mathsf{L}$-shaped nontrivial steady-state regions appear as in the fully connected network case (Fig.~\ref{fig:fully}).
The results of the SFN cases in general show the same $\mathsf{L}$-shaped nontrivial steady state in Figs.~\ref{fig:gamma}(a), \ref{fig:gamma}(b), \ref{fig:gamma}(d), and \ref{fig:gamma}(e) except for Figs.~\ref{fig:gamma}(c) and \ref{fig:gamma}(f) for the positive correlation between the degree and the authority score. We explore such a possibility of nontrivial steady states with $\alpha > 0$ and $p > 0$ values in the next section.  

\subsection{The SFN with positive correlation between authority scores and degrees}
\label{sec:SNF_positive_correlatation}

The nontrivial steady state in the positive correlation case when $\gamma = 2$ in Fig.~\ref{fig:gamma}(c) confirms the existence of the nontrivial steady state ($0 < m(\tau_s) < 1$) with the positive correlation. It confirms that the positive correlation between the heterogeneous network structures and the authority scores effectively blocks the spreading of the opposing opinion with the authoritarian suppression. 
Considering the parameter $p$ forces the nodes to be biased toward the dissenting opinion, it is clear that the positively correlated degrees and authority scores makes the spreading difficult. 
Since the probability $p$ represents the willingness to accept a neighbor's dissenting opinion against the authority, the increment of $m(\tau_s)$ with it in Fig.~\ref{fig:gamma}(c) reflects the crucial role of inner motivation $p$ in the spreading of the dissent opinion. Without it (when $p=0$), the system can be dominated by the affirmative opinion.

We can understand the nontriviality of the positive correlation with the analytic approach in the following.
For simplicity, we assume the completely positively correlated case, i.e., the case that the authority and degree coincide (or at least they are described by the same power-law exponent as mentioned in Sec.~\ref{sec:model}). In addition, 
$p(s_j) = (\gamma - 1) s_j^{-\gamma}$ for the uncorrelated network should be replaced with $p(s_j) = (\gamma - 2) s_j^{1-\gamma}$
because the probability of being a neighbor will be proportional to the neighbor's authority (= degree) and
the exponent for the power-law distribution is modified by $1$ (the celebrated ``friendship paradox''~\cite{Eom2014,Hang2014}).
In that case, the probability becomes
\begin{equation}
\mathrm{Pr}[ q \ge 0 ] = 
\begin{cases}
\displaystyle \frac{\gamma - 2}{2\gamma - 3} \left( \frac{\alpha}{1-\alpha} \right)^{\gamma-1} , & \textrm{for $\alpha \le 1/2$} \\
\displaystyle 1 - \frac{\gamma - 1}{2\gamma - 3} \left( \frac{1-\alpha}{\alpha} \right)^{\gamma-2} , & \textrm{for $\alpha > 1/2$} \,.
\end{cases}
\label{eq:general_gamma_alpha_correlated}
\end{equation}
One can also check the continuity of Eq.~(\ref{eq:general_gamma_alpha_correlated}) at $\alpha = 1/2$. Figure~\ref{fig:Pr}(b) shows the functional form of $\mathrm{Pr}[ q \ge 0 ]$ given by Eq.~(\ref{eq:general_gamma_alpha_correlated}). The stability condition for $0 < m < 1$ requiring $\alpha = 0$ or $p = 0$, therefore, is not affected by the correlated network,
as long as $\gamma > 2$. 
	
For $\gamma \le 2$, things get tricky as the distribution $p(s_j)$ itself cannot be properly normalized (so 
we will need an extra cutoff, such as an exponential tail).
For instance, as $\gamma \to 2^{+}$, $\mathrm{Pr}[ q \ge 0 ] \to 0$ 
according to Eq.~(\ref{eq:general_gamma_alpha_correlated})
[$\gamma = 2.01$ in Fig.~\ref{fig:Pr}(b)],
implying that the $0 < m < 1$ stable state is possible for any $\alpha (< 1)$ and $p$ values. 
This illustrates the situation that spreading of the dissenting opinion can be severely suppressed by dominating hubs (with large degree and authority at the same time) as we confirm with Fig.~\ref{fig:gamma}(c).

Figure~\ref{fig:gamma}(c) also displays the crucial role of the confidence level $\alpha$ as a leading factor for the transmission of the dissent opinion. 
Specifically, $\alpha$ should be larger than a certain threshold $\alpha_{\max,\gamma}$ [as shown in Figs.~\ref{fig:gamma}(c) and \ref{fig:gamma}(f)], e.g., $\alpha_{\max,\gamma = 2} \simeq 0.6$, and $\alpha_{\max,\gamma = 3} \simeq 0.5$.
When $\alpha$ is very small ($\alpha \lesssim 0.1$), the system has a barrier that prevents reaching the consensus of the opposing opinion [see Figs.~\ref{fig:gamma}(c) and \ref{fig:gamma}(f), and Eq.~(\ref{eq:general_gamma_alpha_correlated})]. 
With $\gamma = 2$,  $\mathrm{Pr}[q \ge 0] = 0$ in Eq.~(\ref{eq:general_gamma_alpha_correlated}), so the system is able to deliver the dissenting opinion to the entire system more easily, compared to the case of $\gamma = 3$ [Fig.~\ref{fig:Pr}(c) versus Fig.~\ref{fig:Pr}(f)] where $\mathrm{Pr}[q \ge 0] > 0$ in Eq.~(\ref{eq:general_gamma_alpha_correlated}). 
Only when $\alpha \gtrsim \alpha_{\max,\gamma}$, the successful spreading of the dissenting opinion is possible.  
The point is closely related to the segregated opinion spreading groups that will be discussed later in Sec.~\ref{sec:structure}.
Another notable thing is that the heterogeneous degree distribution, $\gamma = 2$ in this case, with the positive correlation requires a higher confidence level for the spreading of the opposing opinion than the $\gamma = 3$ case, i.e., $\alpha_{\max,\gamma = 2} > \alpha_{\max,\gamma = 3}$.

\subsection{The SFN with negative correlation between authority scores and degrees}
\label{sec:SNF_negative_correlatation}

For the negative correlation, let us consider the case $s_i \propto 1/k_i$ where $k_i$ is the degree of node $i$.
Then, $p(s_j) = \gamma s_j^{-1-\gamma}$ (the probability of being chosen as one's neighbor is inversely proportional to 
the neighbor's authority), which gives
\begin{equation}
\mathrm{Pr}[ q \ge 0 ] = 
\begin{cases}
\displaystyle \frac{\gamma}{2\gamma - 1} \left( \frac{\alpha}{1-\alpha} \right)^{\gamma-1} , & \textrm{for $\alpha \le 1/2$} \\
\displaystyle 1 - \frac{\gamma - 1}{2\gamma - 1} \left( \frac{1-\alpha}{\alpha} \right)^{\gamma} , & \textrm{for $\alpha > 1/2$} \,.
\end{cases}
\label{eq:general_gamma_alpha_negative_correlated}
\end{equation}
In this case, $\mathrm{Pr}[ q \ge 0 ] > 1/2$ for $\alpha = 1/2$, implying that ``your neighbor is weaker than you'' (the ``inverse'' friendship paradox~\cite{Eom2014,Hang2014}).
Figure~\ref{fig:Pr}(c) shows the functional form of $\mathrm{Pr}[ q \ge 0 ]$ given by Eq.~(\ref{eq:general_gamma_alpha_negative_correlated}).
Unlike the positive correlation case, $\mathrm{Pr}[ q \ge 0 ] = 0$ only at $\alpha = 0$ as $\gamma > 1$, so the stability condition is the same as the uncorrelated network case ($\alpha = 0$ or $p = 0$ for the $0 < m < 1$ stability) as shown in Figs.~\ref{fig:gamma}(a) and \ref{fig:gamma}(d).

So far, we have shown that $\gamma \le 2$ with the positive correlation is the only possible nontrivial steady state condition regardless of $\alpha$ and $p$. In other correlations, $\alpha = 0$ or $p = 0$ is the only possible case allowing nontrivial steady states for $\gamma >2$. We also find that finite-size effects are more severe for $\alpha \ll 1$ with large $\gamma$, where the $\alpha$ value small enough to make $\mathrm{Pr}[ q \ge 0 ] \simeq 0$,
which is indeed observable from the results in Fig.~\ref{fig:gamma}: a wider stripe near the horizontal axis.
Note that the assumption $s_i \propto 1/k_i$ is technically different from our negative correlation case in Sec.~\ref{sec:model},
where we just use the inverse order between $\{ s_i \}$ and $\{ k_i \}$. However, we believe that the stability condition will be the same,
based on the robustness of the condition from the uncorrelated to inversely correlated cases described in this section.

\subsection{Effects of underlying network topology}
\label{sec:structure}

\begin{figure}
\begin{center}
\begin{tabular}{lll}
(a)&(b)&(c)\\	\includegraphics[width=0.3\textwidth]{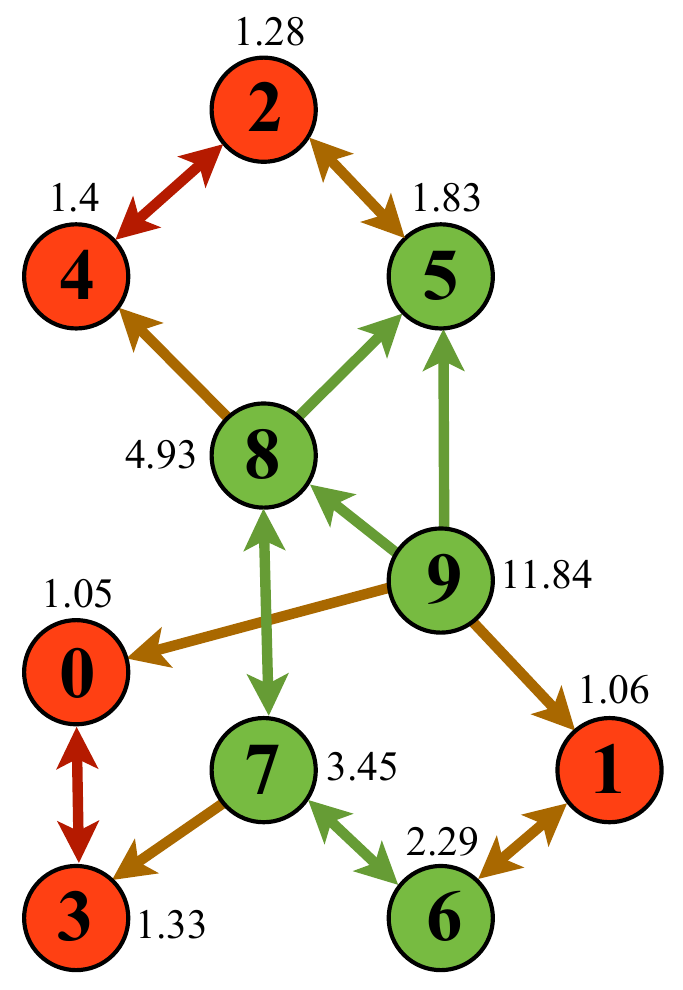}& \includegraphics[width=0.3\textwidth]{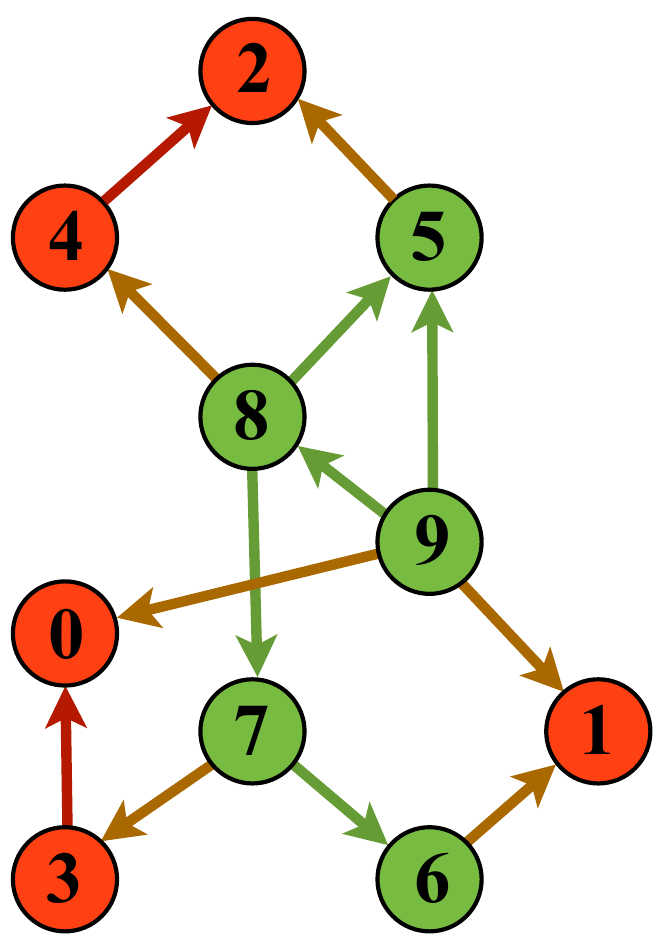}& \includegraphics[width=0.3\textwidth]{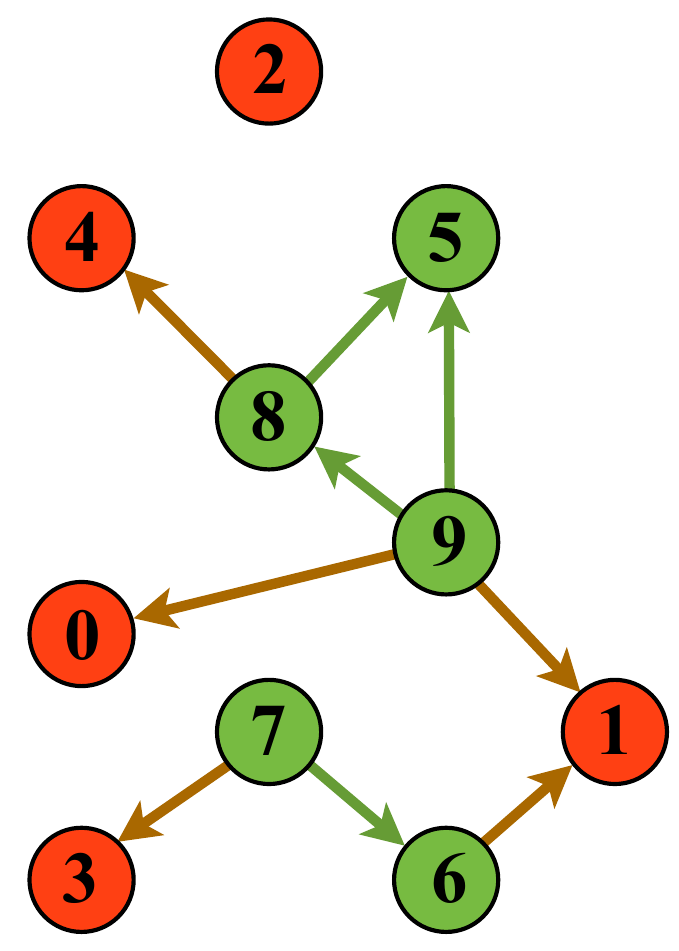}\\
\end{tabular}\quad
\caption{(a) An example of the followship structure representing the comparison process. The red (green) nodes correspond to the agents with the opinion $1$ ($0$), respectively. The red (green) edges connect the two agents with the same opinion $= 1$ ($= 0$), respectively, and the brown edges connect the two agents with different opinions. The authority rank inside the nodes and the actual authority scores outside the nodes are denoted as in Fig.~\ref{fig:model_des}. The followship structure is shown for (b) $\alpha = 0.3$ and (c) $\alpha = 0.6$.}
\label{fig:hidden}
\end{center}
\end{figure}

For a closer examination of the microscopic dynamics of opinion spreading, we focus on the directionality of opinion adoption represented by the directed followship network on top of the undirected substrate network, as illustrated in Fig.~\ref{fig:hidden}. The topological change of the followship network is caused by the sign change of $q$ in Eq.~(\ref{eq:impact}) which is a function of $\alpha$. Consider nodes $2$ and $5$ whose authority scores $s_2 = 1.28$ and $s_5 = 1.83$, respectively, in Fig.~\ref{fig:hidden}(a). As $q(s_2,s_5;\alpha=0.3) < 0$ and $q(s_5,s_2;\alpha=0.3) < 0$ from Eq.~(\ref{eq:impact}), the opinion can spread in both directions ($\textrm{node } 2 \to \textrm{node }5$ and $\textrm{node } 5 \to \textrm{node } 2$) for $\alpha = 0.3$, represented as a bidirectional edge between the two nodes in Fig.~\ref{fig:hidden}(b). In contrast, $q(s_2,s_5;\alpha=0.6) > 0$ and $q(s_5,s_2;\alpha=0.6) > 0$, so no opinion can spread between the two nodes in any direction for $\alpha = 0.6$, represented as the absence of an edge between the two nodes in Fig.~\ref{fig:hidden}(c). Note that it is also possible for the dissenting opinion to spread from node $i$ to $j$ even if $i$ and $j$ are not connected in the followship network (as long as $i$ and $j$ are connected in the original network), since there exists the adoption of a neighbor's dissenting opinion regardless of $q$ (the ``$p$'' process in Fig.~\ref{fig:model_des}), with the probability $p$. 
    
\begin{figure}
\begin{center}
\includegraphics[width=\textwidth]{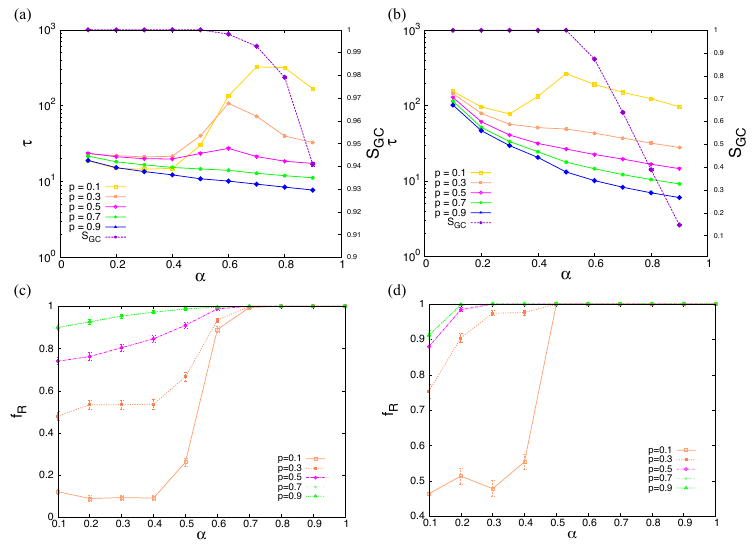}
\end{center}
\caption{The consensus time $\tau$ (the symbols with the solid lines) and the fraction of giant component $S_\mathrm{GC}$ (the symbols with the dashed lines), averaged over $n = 2000$ realizations, are presented in the upper panels. Panel (a) is for $\gamma = 2$, and panel (b) is for $\gamma = 3$. Panels (c) and (d) display the fraction of realizations with the consensus to the dissenting opinion $f_R$ as a function of $\alpha$. Panel (c) is for $\gamma = 2$ and panel (d) is for $\gamma = 3$. The fraction $f_R$ is plotted only when the system reaches to the consensus by $t_\mathrm{max}$ among the $2000$ ensembles ($N=1000$). The lines are the guide to the eyes. The symbols are located at the mean values, and the corresponding error bars represent the standard error of the mean (the error bars of $\tau$ are smaller than the symbols).}
\label{fig:tau_opt}
\end{figure}

This simple example shows that a disjoint group from the giant component (GC) of such a directed followship network can appear depending on the $\alpha$ value, which corresponds to the percolation transition occurring somewhere in between Figs.~\ref{fig:hidden}(b) and \ref{fig:hidden}(c). In that example, node $2$ becomes isolated on the authority comparison level, so there is no way to change its opinion to the dissenting opinion by the comparison process. Only the adoption process of a neighbor's dissenting opinion regardless of the authority scores with the probability $p$ allows the opinion change of this isolated node. In this way, the transmission of the dissenting opinion is related to the structural change for the comparison process.
Figure~\ref{fig:tau_opt} displays the effect of GC on the consensus time $\tau$ for different $\alpha$ and $\gamma$ values. We use the fraction of nodes in the GC, denoted by $S_\mathrm{GC} = n_\mathrm{GC} / n$ where $n_\mathrm{GC}$ is the number of nodes in the giant component and $n$ is the total number of nodes. For simplicity, we neglect the directionality for the connected component analysis, or in other words, we take the weakly connected component. Even though the overall scales of $S_\mathrm{GC}$ are different between Figs.~\ref{fig:tau_opt}(a) and \ref{fig:tau_opt}(b), the location of $\alpha_{\max,\gamma}$ (the $\alpha$ value when the maximum value of $\tau$ occurs for given $\gamma$) marks the segregation of the cluster in $p \leq 0.5$ except for $p=0.1$ at $\gamma =2$ ($\alpha_{\max, \gamma = 2} \simeq 0.6$ and $\alpha_{\max, \gamma = 3} \simeq 0.5$). 

In general, the consensus time $\tau$ increases with $\alpha$ until it reaches the maximum point $\tau_{\max}$ at $\alpha = \alpha_{\max, \gamma}$, and it starts to decrease again when $\alpha > \alpha_{\max, \gamma}$. The result implies that the segregation of opinion-spreading groups affects the dynamics considerably. The sharp increase of $\tau$ near $\alpha = \alpha_{\max,\gamma}$ stems from the fragmented components in the opinion-spreading network, while the decrease of $\tau$ for $\alpha > \alpha_{\max, \gamma}$ is the effect of increment on the viability of the dissenting opinion by the increased confidence parameter $\alpha$. As a substantial fraction of the dissenting opinion survives, the consensus can speed up when $\alpha > \alpha_{\max,\gamma}$ overcoming the authoritarian force suppressing the dissenting opinion. This trend of $\tau$ is similar for both $\gamma = 2$ and $\gamma = 3$ cases, but the reduction of $\tau$ and the prevalence of the dissenting opinion for the $\gamma = 2$ case is much more prominent than that for $\gamma=3$. The exceptional result of $\tau$ at $p=0.1$ for $\gamma = 2$ might come from the relatively large size of GC with small $p \simeq 0.1$. As we explained before, the obvious functional segregation should protect the initial opinion more in the comparison process. However, a number of beholders of the dissenting opinion still has to face the social comparison process due to the small value of $p$, so the system experiences large fluctuations until it reaches the consensus and consumes more time. Nevertheless, the maximum time for the consensus $\tau_\mathrm{max}$ occurs at $\alpha = \alpha_{\max,\gamma=2}$ by the structural segregation in the comparison process.  

So far, we have seen the segregation in the opinion spreading depending on $\alpha$ in terms of consensus time, but it also affects the type of a consensus (whether $m = 1$ or $m = 0$) manifestly [see Figs.~\ref{fig:tau_opt}(c) and \ref{fig:tau_opt}(d)].
In particular, we focus on the fraction of consensus to the dissenting opinion, denoted by $f_R$. As we can observe in Figs.~\ref{fig:tau_opt}(c) and \ref{fig:tau_opt}(d), a drastic increase of $f_R$ occurs at $\alpha_{\max, \gamma=2} \simeq 0.6$ and $\alpha_{\max, \gamma=3} \simeq 0.5$, and $f_R \simeq 1$ for $\alpha \gtrsim \alpha_{\max, \gamma}$. It is caused by the increased viability of the dissenting opinion, as we have explained so far. Given the condition, the main factor responsible for the successful spreading of the dissenting opinion is the existence of isolated groups of agents in the followship network and the type of opinion they protect. As shown in Fig.~\ref{fig:hidden}(c), the separated nodes are generally less influential, so they are initially likely to have the dissenting opinion. Therefore, the isolation can effectively protect the dissenting opinion during the comparison process. We also confirm it by measuring the probability of isolation for each node.  

In summary, the system reaches the opinion consensus more rapidly in the $\gamma = 2$ case which is a more heterogeneous system regarding degree and authority. On the contrary, a more homogeneous structure ($\gamma =3$) promotes the spreading of the dissenting opinion more, even though it takes longer time, compared to the $\gamma=2$ case. For the spreading of the dissenting opinion, a confidence level exceeding a certain threshold confidence level is required, i.e., $\alpha > \alpha_{\max,\gamma}$, and it can lead the successful spreading of the dissenting opinion with low values of $p < 0.1$. 

\section{Summary and conclusions}

We have introduced a stylized opinion formation model to understand the spreading of the less influential agents' opinion by setting the correlation between the degree and the authority score, with personal characteristics such as the confidence level and the willingness to accept the dissenting opinion. First of all, nonzero amount of willingness to accept the dissenting opinion, albeit small, drastically promotes the spreading of the dissenting opinion in most cases. Nontrivial steady states, or the coexistence of different opinions, is possible when the degree and the authority score are positively correlated in the case of severe heterogeneity, namely, when the power-law exponent $\gamma \le 2$. It implies that the strong authoritarian structure, combined with its correlation to the number of neighbors, efficiently suppresses the acceptance of the opposing opinion from less influential people.

For given heterogeneity and correlation, the confidence level $\alpha$ toward the agents' own opinion is the major factor deciding the prevalence of the dissenting opinion, even though it may take a long time. In particular, the parameter $\alpha$ controls the viability of the dissenting opinion, via the segregation of the opinion-spreading subgroups in the comparison process. By incorporating the personal factors in the model, we have learned two things. First, if a population has the willingness to hear the dissenting opinion held by less influential people, there is a large chance for the effective spreading of their opinion. Second, from the critical role of the confidence level, we infer that if the confidence of each agent is strong enough, the opinion of less influential people can be spread to the entire population even in the case of severely heterogeneous authority and degree distribution.  
Since our model is a highly simplified one, it may not capture all of the details in the real opinion formation processes in our society. For example, the assumption of the same power-law exponent for the authority and degree distribution could be a limitation of the model. In spite of the limitations, though, the results suggest a possibility of the crucial impact on the correlations between the network and authority structures in opinion formation. For future studies, with the necessity of studies with a spreading of misinformation~\cite{pnas1,pnas2} from the authorities, the model could be applied to figure out effects of the authority and a correlated network on the spreading of misinformation in a hierarchical structure. 

\section*{Acknowledgements}
This work was supported by the Human Resources Development program (No.~20124010203270) of the Korea Institute of Energy Technology Evaluation and Planning (KETEP) grant funded by the Korea government Ministry of Trade, Industry and Energy. We thank Heetae Kim, Minjin Lee, Hang-Hyun Jo, and Jinhyuk Yun for helpful discussions and suggestions, and Jimin Hwang for linguistic advice. Computation was partially carried out using a server in Complex Systems and Statistical Physics Lab, Korea Advanced Institute of Science and Technology.









\end{document}